\documentclass[global,twocolumn]{svjour}
\usepackage{graphicx}
\usepackage{amssymb}
\usepackage{amsmath}
\usepackage{hyperref}
\usepackage{graphics}
\hypersetup{pdfpagemode=UseNone,
    bookmarks=false,         % show bookmarks bar?
    unicode=false,          % non-Latin characters in Acrobats bookmarks
    pdftoolbar=false,        % show Acrobats toolbar?
    pdfmenubar=true,        % show Acrobats menu?
    pdffitwindow=true,      % page fit to window when opened
    colorlinks=true,       % false: boxed links; true: colored links
    linkcolor=blue,          % color of internal links
    filecolor=black,      % color of file links
    urlcolor=blue,           % color of external links
}

\journalname{Appl. Phys. B}
\begin{document}

\title{Monolayer graphene as dissipative membrane in an optical resonator}

\author{Hendrik M. Meyer, and Moritz Breyer, and Michael K{\"o}hl}

\institute{Physikalisches Institut, University of Bonn, Wegelerstra{\ss}e 8, D-53115 Bonn, Germany \email{hmeyer@physik.uni-bonn.de}}

\date{\today}
\maketitle

\begin{abstract}
We experimentally demonstrate coupling of an atomically thin, free-standing graphene membrane to an optical cavity.  By changing the position of the membrane along the standing-wave field of the cavity we  tailor the dissipative coupling between the membrane and the cavity, and we show  that the dissipative coupling can outweigh the dispersive coupling. Such a system, for which controlled dissipation prevails dispersion, will prove useful for novel laser-cooling schemes in optomechanics. In addition, we have determined the continuous-wave optical damage threshold of free-standing monolayer graphene of 1.8(4)~MW/cm$^2$ at 780nm. 
\end{abstract}

%\section{Introduction}

%Zitate eingefügt
The coupling between light and matter is a fundamental ingredient for many applications ranging from highly precise detection of mechanical motion \cite{anetsberger2009near,favero2009optomechanics} to quantum information science \cite{kimble2008quantum}. Accurate control over the amplitude and phase of the coupling  requires precise localization of the matter relative to the light field. Such control has been experimentally demonstrated using microscopic, point-like physical systems such as single atoms \cite{lee2014three}, trapped ions \cite{guthohrlein2001single,steiner2013single}, gold nanoparticles \cite{mader2015scanning}, semiconductor quantum dots \cite{Sanchez2013} and NV centers in diamond \cite{Albrecht2013}. Additionally, macroscopic objects, such as SiN membranes have been coupled locally to standing-wave light fields \cite{zwickl2007high,jayich2008dispersive}. They combine low absorption and high dispersion with very good handling, and they have been used, for example, in optomechanical experiments \cite{Aspelmeyer2014} in which the motion-dependent interaction between the membrane and a light field is studied. 

Advances in the fabrication of two-dimensional materials, like graphene \cite{meyer2007structure}, have made it possible to fabricate atomically thin membranes, which combine the ease of use of macroscopic membranes with the positioning accuracy of a single atom. Owing to their low mass and high stiffness they promise higher vibrational frequencies  \cite{chen2009performance,singh2014optomechanical} than SiN membranes, which is of great interest for future graphene--based optomechanical applications. Another key difference between membranes made of graphene and those made of SiN is the nature of the light-matter coupling. While for SiN membranes the coupling is  dispersive, for  graphene membranes it has been predicted to be mostly dissipative since a monolayer of graphene exhibits a single--pass absorption of $A \approx \pi \alpha = 2.3\%$ in the optical range \cite{nair2008fine}. Hence, a graphene membrane will cause a  position-dependent dissipation of the intra-cavity field, which is wavelength-independent within the visible to near infra-red spectral range. In contrast to standard dispersive optomechanical coupling \cite{Aspelmeyer2014}, a predominantly dissipative coupling between membrane and cavity field could allow for efficient laser cooling of the membrane's motion even outside the resolved sideband regime \cite{PhysRevLett.102.207209,PhysRevLett.107.213604}.    

Recently, graphene has been optomechanically coupled to superconducting microwave cavities \cite{singh2014optomechanical} and to the evanescent field of an optical microsphere-resonator \cite{PhysRevApplied.3.024004}, however, in both cases the spatial structure of the electromagnetic field was not resolved in the coupling. Here we present an experimental study of a free-standing layer of graphene inside a Fabry-Perot resonator. We demonstrate that the position of the membrane with respect to the cavity standing-wave field leads to a strong and highly controlled modulation of the cavity losses due to dissipation. Furthermore, we find that the change of the cavity linewidth due to dissipation is three times larger than the dispersive frequency shift caused by the membrane, which realizes for the first time a regime dominated by dissipation rather than dispersion. 

%Figure 1 - Experimental Setup.
\begin{figure}
\includegraphics[width=\columnwidth]{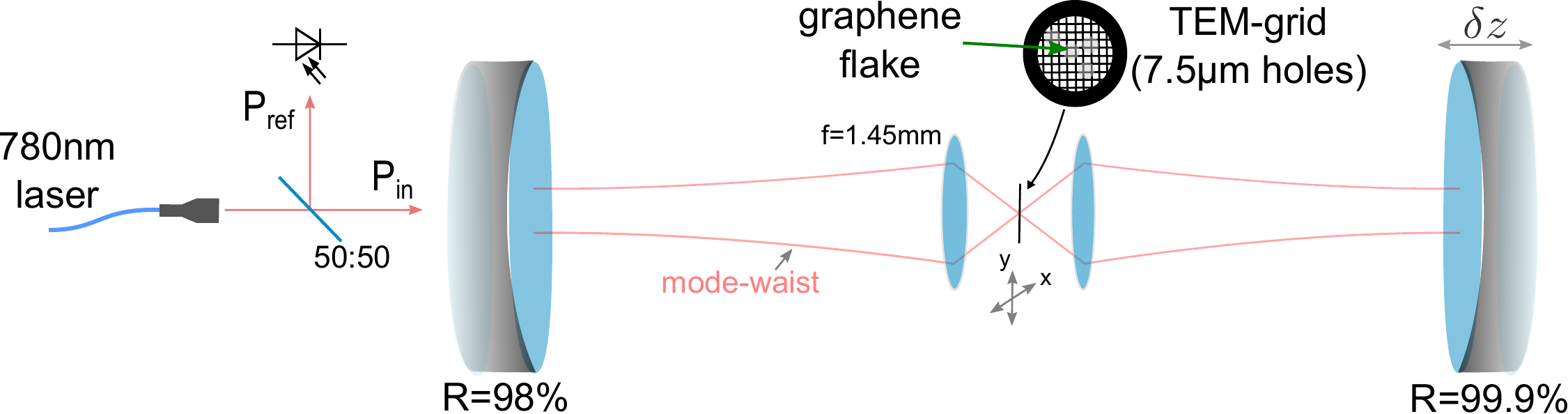}%
\caption{(Color online) An optical Fabry-Perot resonator is loaded with a free-standing monolayer of graphene. The position of the graphene sample can be fine--tuned in the x-y-plane using a piezo stage. Light is coupled into the cavity from the low-reflectivity side and the reflection from the cavity is monitored using a 50:50 beamsplitter and a photodiode. The high-reflectivity mirror is mounted on a piezo transducer in order to adjust the cavity length.}
\label{setup}
\end{figure}
%
%Figure 2 - standing wave.
\begin{figure*}[t]
\includegraphics[width=\textwidth]{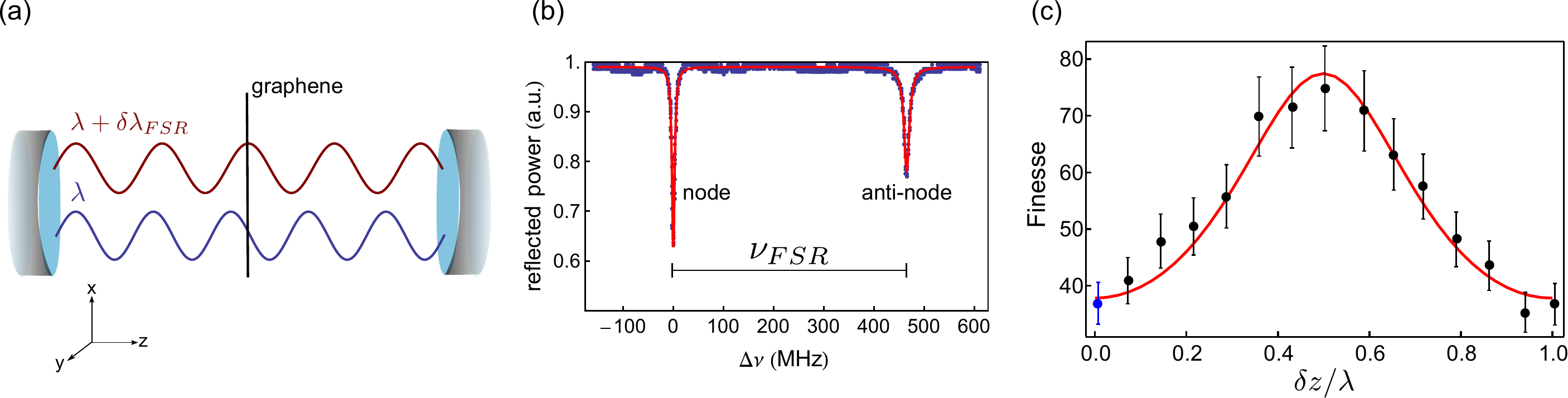}%
\caption{\label{standingwave}(Color online) (a) Localization of a layer of graphene with respect to the cavity standing-wave field. When the frequency of the laser is changed by one free spectral range of the cavity, the single layer shifts from a node into an anti-node of the standing-wave pattern. (b) Reflected power from the cavity while scanning the laser across one free spectral range. On the left resonance graphene is located at a node, while on the right resonance it is located at an anti-node. (c) Finesse of the cavity for different relative positions of the graphene layer and the anti-node of the standing-wave field. The point at $\delta z/\lambda = 0$ (blue) is extracted from the data shown in (b). The solid line is derived from the numerical model discussed in the main text.}
\end{figure*}
Our experimental setup is shown in Fig. 1. It is composed of a Fabry-Perot optical resonator with a free-standing layer of graphene in the ``membrane-in-the-middle'' configuration. The position of the graphene with respect to the standing wave can be experimentally controlled which allows us to influence the dissipation and hence the cavity loss rate. The latter is reflected by an increased linewidth and a reduced in-coupling into the cavity when the membrane is moved from a node to an anti-node of the standing-wave field. In order to achieve a good coupling between the cavity mode and the graphene layer, the mode waist of the cavity must be much smaller than the physical size of free-standing graphene flakes, which typically is on the order of  $10\mu$m. 

In order to address  the challenge of working with small free-standing samples, we have designed an optical cavity featuring two integrated lenses to achieve an experimentally verified mode-waist of $w_0=(1.2\pm 0.2)\mu \text{m}$. The cavity has a total length of $l_{\text{cav}}=32$~cm, corresponding to a free spectral range of $\nu_{\text{FSR}}=c/(2l_{\text{cav}})=470$~MHz (see \autoref{setup}). We perform reflection spectroscopy on the cavity  using light from a single-mode external-cavity diode laser at 780~nm. For reference, we first measure the finesse of the empty cavity to be  $\mathcal{F}_e=2\pi/\mathcal{L}_e=\nu_{\text{FSR}}/\delta \nu_e = 83\pm 9$, where $\delta \nu_e =(5.6 \pm 0.6)~\text{MHz}$ is the linewidth of the cavity and $\mathcal{L}_e=(7.5\pm0.8)\%$ are the  total losses. Both the transmission of the mirrors and the intra-cavity optics contribute to the losses. The linewidth of the cavity determines the electric field decay rate by $\kappa_e=2\pi\times\delta \nu_e/2=2\pi\times(2.8\pm 0.3)~\text{MHz}$. Next, we introduce a free-standing graphene sample into the Fabry-Perot resonator. We use a commercially available sample grown by chemical vapor deposition (CVD) and transferred onto a transmission electron microscopy (TEM)~-grid \cite{sample}. The graphene is mostly a single-layer sample but also a few two-layer flakes have been detected, see below.

%Experimental Oberservation of the modulation:
Initially, we position the graphene sample at the node of the standing-wave field by minimizing the losses (linewidth) of the cavity. When subsequently sweeping the center frequency of the laser by one free spectral range of the cavity ($\Delta \nu_L=\nu_{FSR}$), the relative position between graphene and the standing-wave changes from a node to an anti-node as depicted in \autoref{standingwave} (a). Resultingly, the losses of the resonator are different for the two laser frequencies  and allow us to perform a differential measurement of the cavity losses caused by the graphene layer without the need for mechanical realignment. An example of such a measurement is shown in \autoref{standingwave} (b). We can also access other locations of the graphene sample relative to the standing wave by changing the cavity length by $\delta z$ using a piezo transducer at one of the cavity mirrors and perform a similar measurement. In \autoref{standingwave} (c) we show the complete results of how the cavity finesse changes when positioning the graphene sample at different locations within the standing wave.  The results highlight that the periodic structure of the cavity losses can be clearly resolved. The observed periodicity of $\lambda$ stems from moving the cavity mirror rather than the graphene sample.

We fit our experimental data with a theoretical model which computes the total cavity dispersion and losses from the dispersion and absorption of each element in the resonator and then solves for the condition of a stable resonator round trip \cite{jayich2008dispersive}. In the model we have used the reflection and transmission amplitudes for graphene calculated from the Fresnel formulas for a  membrane of thickness $d_g=0.345~\text{nm}$ and with a complex index of refraction of $n_g = 2.71 + 1.41~i$ \cite{cheon2014reliably}. Together with the empty cavity loss of $\mathcal{L}_e =8\%$, which is in agreement with the data for the graphene at the node, this model fits the data well, as can be seen from the solid line in Figure 2c. For a single layer of graphene we find that the decay rate of the cavity electric field  increases by $\Delta \kappa  = 2\pi\times(3.5 \pm 0.4)~\text{MHz}$, which quantifies the dissipative coupling between the graphene flake and the resonator mode. %The comparison to the decay rate of the empty cavity shows that these are on the same order since $\Delta \kappa/\kappa_e \approx 1.2$.

On different sections of our graphene sample, we have have observed different values of the absorption. In particular, there are three different discrete values of the finesse, which we attribute to different number of layers. In order to analyze this effect, we extend the numerical model to analyze how many layers of graphene contribute to the absorption. In Figure 3, the dashed line shows the behavior of the cavity finesse as a function of the (continuous) thickness of the absorber. The data points show the measurements and we find that they match the curve equidistantly, and hence count the numbers of layers. The behavior can also be understood in the following simplified way: an absorber with  a single-pass loss $A$ leads to an average loss of $2A$ in the standing-wave cavity. From the $\cos^2(2\pi x/\lambda)$ structure of the standing wave field it follows that the peak loss of a sub-wavelength thick absorber is $4A$. In lowest order we assume that the single-pass loss is proportional to the number $N$ of layers and hence the cavity finesse for the graphene sample at the anti-node behaves as $\mathcal{F}=2\pi/(\mathcal{L}_e+4N\cdot A)$. The squares in Figure 3 display the result of this estimate using $A=2.3\%$, which agrees well with the data.

%Figure 3 - Finesse vs. number of layers
\begin{figure}
\includegraphics[width=\columnwidth]{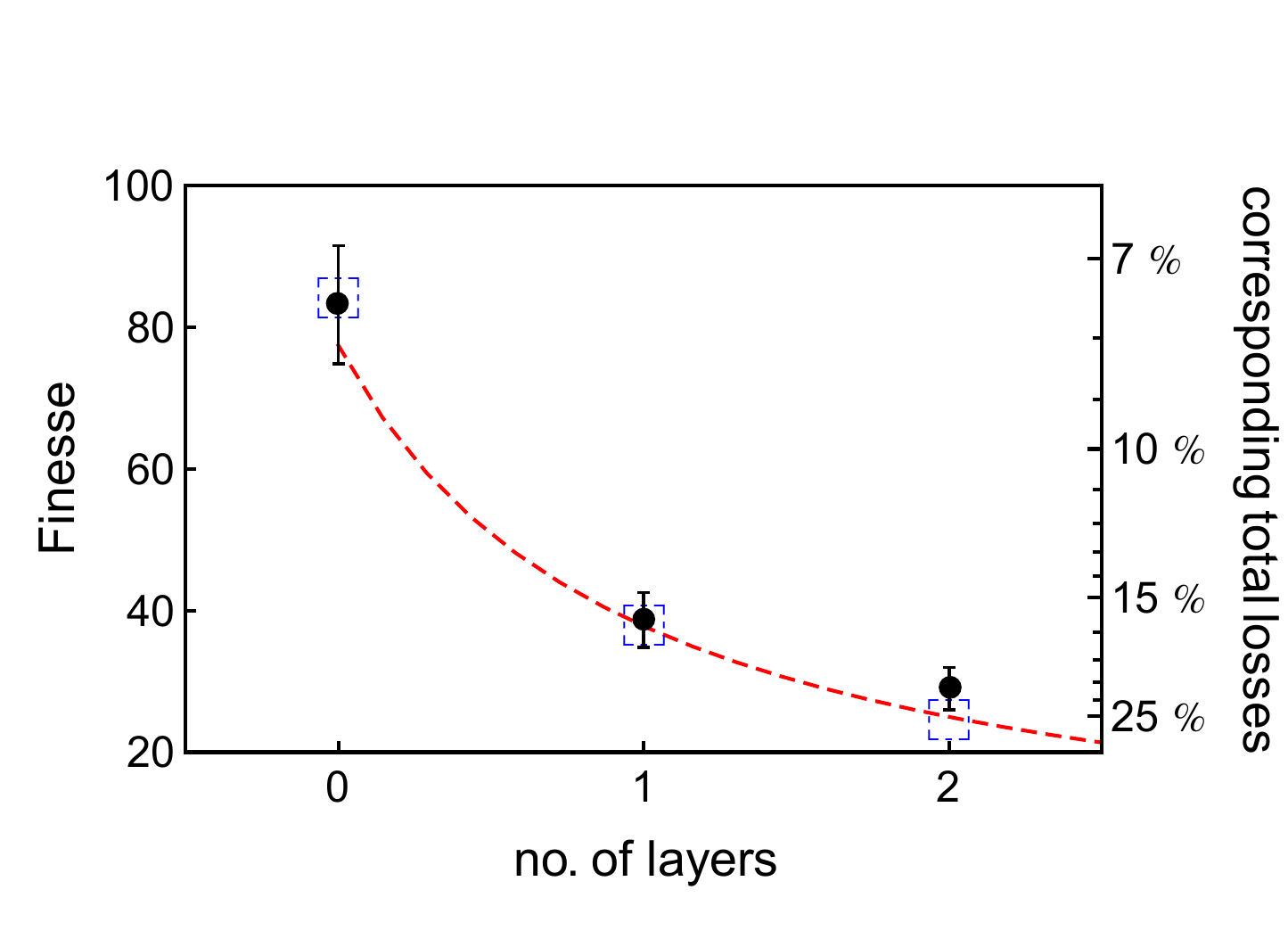}%
\caption{\label{layer}(Color online) Cavity finesse vs. number of graphene layers located at the cavity anti-node. The dashed line is obtained from the simulation with the parameters obtained in \autoref{standingwave} (c). The empty squares (blue) assume a loss of $4N\cdot A$ for $N$ layers. }
\end{figure}
%
%Figure 4 - Damage threshold / Dispersion measurement
\begin{figure}
\includegraphics[width=\columnwidth]{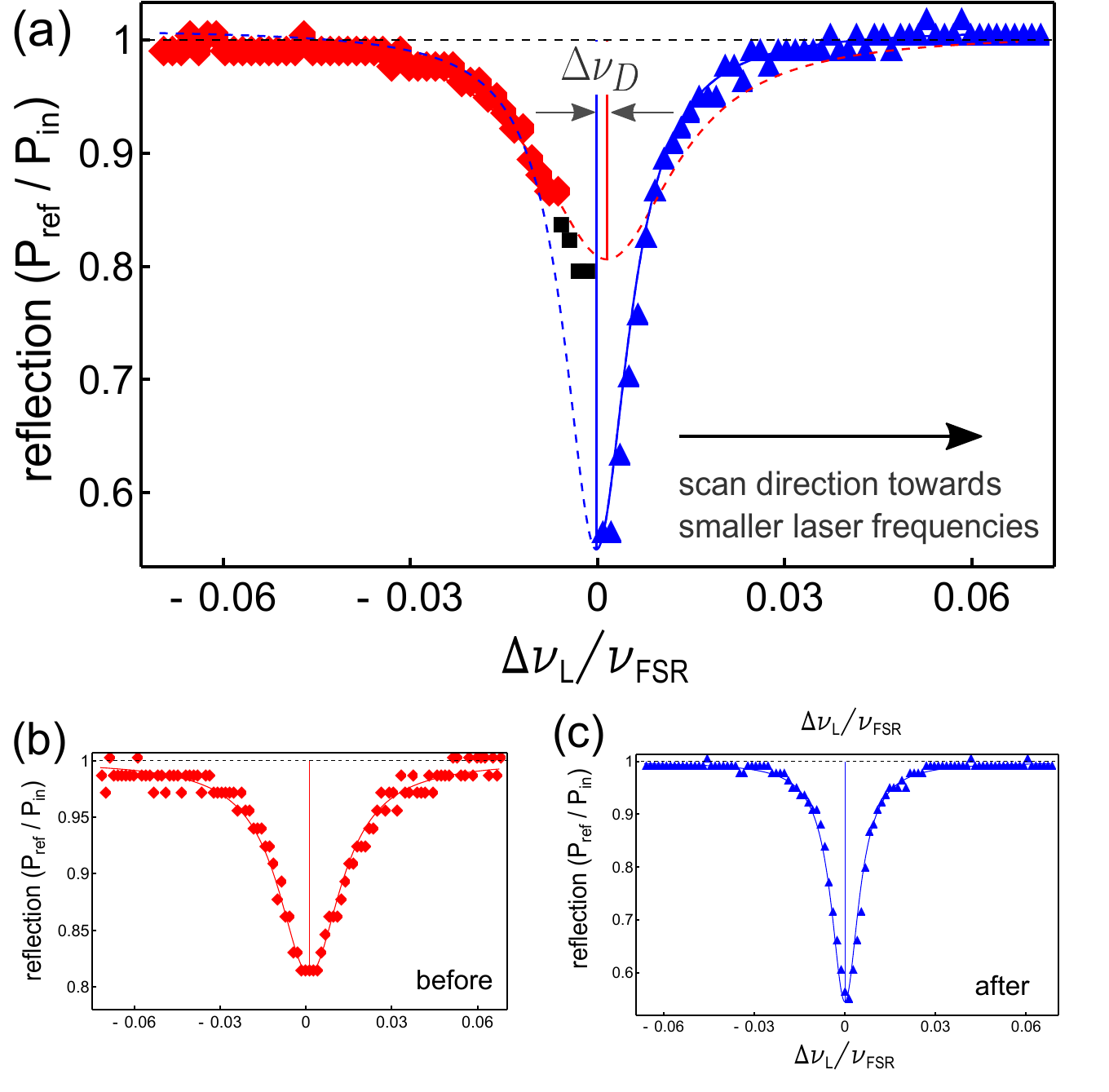}%
\caption{(Color online) Measurement of the dispersive shift caused by a single layer of graphene. (a) Cavity transmission during the frequency scan of the laser at an intensity higher than the damage threshold of graphene. Near the peak coupling into the resonator, the graphene is optically damaged and becomes transparent which leads to a shift of the cavity resonance. Red data points (diamonds): undamaged graphene; blue data points (triangles): damaged graphene. The dispersive shift $\Delta \nu_D$ is determined from Lorentzian line fits to the wings of the data. The black data points near to the threshold have not been used for the fit. Also shown are reference scans at low intensity before (b) and after (c) the dispersive measurement to determine the width and amplitude of the Lorentzian fits.}
\label{damage}
\end{figure}

So far, we have demonstrated that positioning of a graphene membrane at specific positions within the cavity field leads to controlled dissipation. We now set this into relation to the dispersive shift, which is the change of the resonance frequency of the cavity caused by the presence of the graphene layer. This is an important quantity since certain optomechanical laser-cooling schemes require optical dissipation--dominated coupling rather than dispersive coupling between membrane and cavity \cite{PhysRevLett.102.207209,PhysRevLett.107.213604}. However,  membranes with such optical properties have not yet been demonstrated experimentally. Considering the change in the optical path length of the cavity by the graphene membrane, we theoretically estimate the dispersive frequency shift  $\Delta \omega_D = -2\pi \times \nu_{\text{FSR}}~\frac{2d_g}{\lambda}~\left[\operatorname{Re}(n_g)-1\right]=-2\pi \times 0.71~\text{MHz}$. Hence, we expect the dispersive shift to be smaller than the total linewidth of our resonator and towards smaller absolute resonance frequency as the optical path length becomes longer when graphene is placed inside the resonator.
%Experimental Observation:

In order to experimentally measure the dispersive shift without the efforts of homodyne detection, we position the graphene at the  anti-node of the cavity field and  apply a laser power above the optical damage threshold of graphene when resonantly coupled into the resonator. When scanning the laser across the cavity resonance, the graphene layer gets destroyed when the light couples into the resonator. This leads to a quick change of the cavity resonance frequency by the dispersive shift $\Delta \omega_D=2\pi\times\Delta \nu_D$ (see \autoref{damage}). It is important to note that the cavity response time is about $\kappa_e^{-1} \sim 50\,\text{ns}$, which is two orders of magnitude smaller than the separation between the data points and therefore the cavity response is instantaneous. In order to quantify the dispersive shift, the data are divided into two parts representing the loaded and empty cavity (red (diamonds) and blue (triangles) data in Figure 4a). These parts can be identified by a strong change in the cavity in-coupling efficiency. We fit each of the wings with a Lorentzian profile using the linewidth obtained from the previous measurement and determine the line-centers of the Lorentzian. With this method, we find a dispersive shift of $\Delta \omega_D  = -2\pi \times (1.0 \pm 0.4)$~MHz, in agreement with our expectation.

From the dispersive and the absorptive measurements, it is possible to calculate the corresponding optomechanical coupling constants obtained when graphene is placed in the center between node and anti-node by $G\lesssim k\cdot\left|\Delta\omega_D\right| = 2\pi \times (8.3\pm2.5)~\text{kHz/nm}$ and $\Gamma_{dp}=k\cdot\Delta\kappa = 2\pi \times (28.3 \pm 3.3)~\text{kHz/nm}$, where $k=2\pi/\lambda$ is the wavenumber (see supplementary). From this we find a ratio between dissipative and dispersive coupling of 
\begin{equation}
\Gamma_{dp}/G\gtrsim \Delta \kappa/\Delta \omega_D = 3.4 \pm 1.1~,
\label{eqn:intensity}
\end{equation}
which exceeds previous measurements in other systems \cite{PhysRevApplied.3.024004,PhysRevX.4.021052} by approximately one order of magnitude. Moreover, it shows the unusual features of the graphene membrane as compared to standard optomechanical membranes, such as SiN, for which this ratio typically is in the range of $\sim 10^{-4}$.\\

%Taking both the dispersive as well as the absorptive measurements, the ratio of dissipative and dispersive coupling is
%\begin{equation}
%\Delta \kappa/\Delta \omega_D = 3.4 \pm 1.1
%\label{eqn:intensity}
%\end{equation} 
%This result exceeds previous measurements \cite{PhysRevApplied.3.024004,PhysRevX.4.021052} by approximately one order of magnitude. Moreover, it shows the unusual features of the graphene membrane as compared to standard optomechanical membranes, such as SiN, for which this ratio typically is in the range $10^{-4}$.

%Damage Threshold:
From our measurements, we  also deduce the optical cw-damage threshold of graphene for $\lambda=780\,$nm, which is interesting for instance for applications in nonlinear optics  \cite{PhysRevLett.105.097401} or to interface other nanoscopic system like nanodiamonds \cite{brenneis2015ultrafast}. To this end, we calculate the intensity of the cavity field at the anti-node from a numerical model \cite{jayich2008dispersive}. Our data yield a cw-damage threshold of $I_D=(1.8 \pm 0.4)$~MW/cm\textsuperscript{2}, which is the average of ten measurements. This result is comparable to the optical damage threshold of graphene mounted on silicon substrates where lattice modifications caused by bond breaking have been observed \cite{PhysRevB.79.165428}. We note that below the damage threshold intensity, no saturation of the graphene absorption has been observed. This is in contrast to experiments with short (ps) laser pulses, where the damage threshold \cite{Roberts2011} is several orders of magnitude larger than the saturation intensity \cite{graphenefibertip}. 

%--------------------------
%\section{Conclusion and Outlook}
We conclude that a single free-standing monolayer of graphene in a membrane-in-the-middle configuration is a promising experimental platform for cavity optomechanics in the dissipation-dominated regime. Along with the good mechanical properties of graphene, which have been verified by evanescent motional readout \cite{PhysRevApplied.3.024004}, theoretically predicted dissipative ground state cooling schemes \cite{xiao2014dissipative} might become possible in future experiments.

\section*{Appendix}

For optomechanical applications, the coupling constants are of fundamental interest. In general they can be divided into dispersive coupling $G$, which is the change of the cavity resonance frequency $\Delta \omega$ with the membrane displacement along the cavity axis (z-axis) and into dissipative coupling $\Gamma_{dp}$, which is the equivalent change of the cavity electric field decay rate $\Delta \kappa$. In general these coupling constants can be written as \cite{Wu2014}:
\begin{equation}
G(z)=\frac{\partial \omega_0(z)}{\partial z}~,~\Gamma_{dp}(z)=\frac{\partial \kappa(z)}{\partial z}~.
\end{equation}
\begin{figure*}[t]
\includegraphics[width=0.9\textwidth]{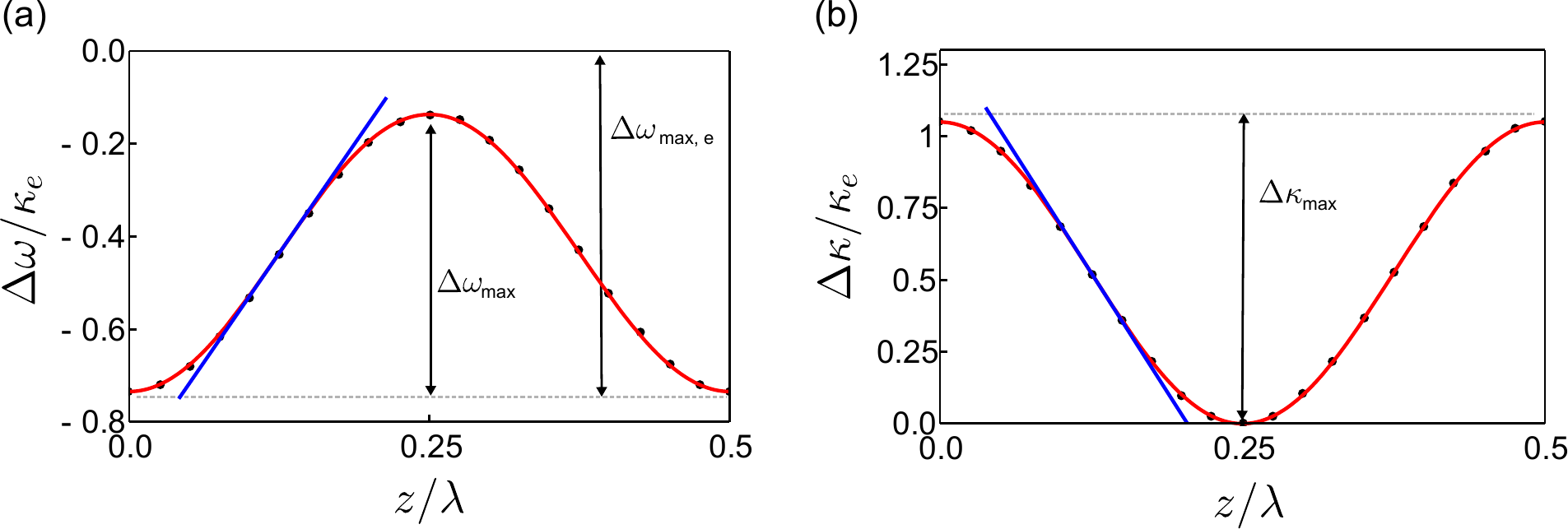}%
\caption{\label{omegakappatheory}(Color online) (a) Change of the resonance frequency $\Delta \omega$ and (b) change of the cavity electric field decay rate $\Delta \kappa$ with respect to the empty cavity when graphene is moved from an anti-node ($z /\lambda = \{0,0.5\}$) to a node ($z /\lambda = 0.25$) of the cavity standing-wave field. Both values are given in units of the empty cavity electric field decay rate $\kappa_e$. The simulation data (black dots) are fitted with a suitable fit-function (red curve, see equation \ref{eq:omegafit}-\ref{eq:kappafit}). The coupling constants reach their maximum in the center between node and anti-node ($z=\lambda/8$), where the linear approximation of the curve is shown (blue line).}
\end{figure*}
Because the electric field along the cavity-axis forms a periodic standing-wave, both of these coupling constants also follow a periodic pattern along the cavity axis. In \autoref{omegakappatheory} the theoretical values of $\Delta \omega = \omega_0-\omega_{0,e}$ and $\Delta \kappa = \kappa - \kappa_e$ are shown, where $\omega_{0,e}$ and $\kappa_e$ denote the resonance frequency and the electric field decay rate of the empty cavity, respectively. Here the same simulation and parameters as given in the main text have been used. Because the length scale along the cavity-axis is set by the wavelength, the slope and therefore the coupling constants are fully defined by the amplitudes $\Delta \omega_{max}$ and $\Delta \kappa_{max}$ when the membrane is moved from a node into an anti-node of the standing-wave field (see \autoref{omegakappatheory}). In order to calculate the coupling constants, we fit the simulation using suitable fit-functions defined by
\begin{align}
\Delta \omega &=\frac{\Delta \omega_{max}}{2} \cos\left(2 k z\right)+y_{0,\omega} \label{eq:omegafit}\\
\Delta \kappa &=\frac{\Delta \kappa_{max}}{2} \cos\left(2 k z\right)+y_{0,\kappa} \label{eq:kappafit}~,
\end{align}
where $k=2\pi/\lambda$ is the wavenumber of the intra-cavity field. From this follows:
\begin{equation}
G=\frac{\partial \omega_{0}(z)}{\partial z} =\frac{\partial \Delta \omega}{\partial z} = k\cdot \Delta \omega_{max} \sin\left(2 kz\right)~ = G_{max}\sin\left(2 kz\right),
\label{eq:G}
\end{equation}
and equivalently $\Gamma_{dp}=k\cdot\Delta\kappa_{max}\sin\left( 2kz\right)= \Gamma_{dp,max}\sin\left( 2kz\right)$. Therefore the absolute values become largest when the membrane is placed in the center between node and anti-node ($z = \lambda/8$), as here the slope of the curves in \autoref{omegakappatheory} is largest. While each individual coupling constant even approaches zero when the membrane is placed close to a node or anti-node of the standing-wave field, the ratio of the coupling constants is independent of the exact location as 
\begin{equation}
\frac{\Gamma_{dp}}{G}=\frac{\Delta \kappa_{max}}{\Delta \omega_{max}}.
\label{eq:ratio}
\end{equation}
In our experiment we measured the shift of the resonance frequency of the cavity when the membrane is removed from an anti-node of the standing-wave field. In \autoref{omegakappatheory} this corresponds to the theoretical value indicated $\Delta \omega_{max,e}$. If we use this value for calculating the coupling constant $G_{max}$ using \autoref{eq:G}, our simulation suggests that we overestimate the dispersive coupling by about 20\%. Therefore generally $\Gamma_{dp}/G \gtrsim \Delta \kappa_{max}/\Delta \omega_{max,e}$, which is the value we quote in equation (1) of the main text.   

%SECTION: 2

%\section{Acknowledgments}
This work has been supported by BCGS and the Alexander-von-Humboldt Stiftung.


\begin{thebibliography}{10}

\bibitem{anetsberger2009near}
G.~Anetsberger, O.~Arcizet, Q.~P. Unterreithmeier, R.~Rivi{\`e}re,
  A.~Schliesser, E.~M. Weig, J.~P. Kotthaus, and T.~J. Kippenberg, ``Near-field
  cavity optomechanics with nanomechanical oscillators,'' {\em Nature Physics},
  vol.~5, no.~12, pp.~909--914, 2009.

\bibitem{favero2009optomechanics}
I.~Favero and K.~Karrai, ``Optomechanics of deformable optical cavities,'' {\em
  Nature Photonics}, vol.~3, no.~4, pp.~201--205, 2009.

\bibitem{kimble2008quantum}
H.~J. Kimble, ``The quantum internet,'' {\em Nature}, vol.~453, no.~7198,
  pp.~1023--1030, 2008.

\bibitem{lee2014three}
M.~Lee, J.~Kim, W.~Seo, H.-G. Hong, Y.~Song, R.~R. Dasari, and K.~An,
  ``Three-dimensional imaging of cavity vacuum with single atoms localized by a
  nanohole array,'' {\em Nature communications}, vol.~5, 2014.

\bibitem{guthohrlein2001single}
G.~Guth{\"o}hrlein, M.~Keller, K.~Hayasaka, W.~Lange, and H.~Walther, ``A
  single ion as a nanoscopic probe of an optical field,'' {\em Nature},
  vol.~414, no.~6859, pp.~49--51, 2001.

\bibitem{steiner2013single}
M.~Steiner, H.~M. Meyer, C.~Deutsch, J.~Reichel, and M.~K{\"o}hl, ``Single ion
  coupled to an optical fiber cavity,'' {\em Physical Review Letters},
  vol.~110, no.~4, p.~043003, 2013.

\bibitem{mader2015scanning}
M.~Mader, J.~Reichel, T.~W. H{\"a}nsch, and D.~Hunger, ``A scanning cavity
  microscope,'' {\em Nature communications}, vol.~6, 2015.

\bibitem{Sanchez2013}
J.~Miguel-Sanchez, A.~Reinhard, E.~Togan, T.~Volz, A.~Imamoglu, B.~Besga,
  J.~Reichel, and J.~Estève, ``Cavity quantum electrodynamics with
  charge-controlled quantum dots coupled to a fiber fabry–perot cavity,''
  {\em New Journal of Physics}, vol.~15, no.~4, p.~045002, 2013.

\bibitem{Albrecht2013}
R.~Albrecht, A.~Bommer, C.~Deutsch, J.~Reichel, and C.~Becher, ``Coupling of a
  single nitrogen-vacancy center in diamond to a fiber-based microcavity,''
  {\em Phys. Rev. Lett.}, vol.~110, p.~243602, Jun 2013.

\bibitem{zwickl2007high}
B.~M. Zwickl, W.~E. Shanks, A.~M. Jayich, C.~Yang, A.~C. Bleszynski~Jayich,
  J.~D. Thompson, and J.~G.~E. Harris, ``High quality mechanical and optical
  properties of commercial silicon nitride membranes,'' {\em Applied Physics
  Letters}, vol.~92, no.~10, 2008.

\bibitem{jayich2008dispersive}
A.~Jayich, J.~Sankey, B.~Zwickl, C.~Yang, J.~Thompson, S.~Girvin, A.~Clerk,
  F.~Marquardt, and J.~Harris, ``Dispersive optomechanics: a membrane inside a
  cavity,'' {\em New Journal of Physics}, vol.~10, no.~9, p.~095008, 2008.

\bibitem{Aspelmeyer2014}
M.~Aspelmeyer, T.~J. Kippenberg, and F.~Marquardt, ``Cavity optomechanics,''
  {\em Rev. Mod. Phys.}, vol.~86, pp.~1391--1452, Dec 2014.

\bibitem{meyer2007structure}
J.~C. Meyer, A.~K. Geim, M.~Katsnelson, K.~Novoselov, T.~Booth, and S.~Roth,
  ``The structure of suspended graphene sheets,'' {\em Nature}, vol.~446,
  no.~7131, pp.~60--63, 2007.

\bibitem{chen2009performance}
C.~Chen, S.~Rosenblatt, K.~I. Bolotin, W.~Kalb, P.~Kim, I.~Kymissis, H.~L.
  Stormer, T.~F. Heinz, and J.~Hone, ``Performance of monolayer graphene
  nanomechanical resonators with electrical readout,'' {\em Nature
  nanotechnology}, vol.~4, no.~12, pp.~861--867, 2009.

\bibitem{singh2014optomechanical}
V.~Singh, S.~Bosman, B.~Schneider, Y.~M. Blanter, A.~Castellanos-Gomez, and
  G.~Steele, ``Optomechanical coupling between a multilayer graphene mechanical
  resonator and a superconducting microwave cavity,'' {\em Nature
  nanotechnology}, vol.~9, no.~10, pp.~820--824, 2014.

\bibitem{nair2008fine}
R.~Nair, P.~Blake, A.~Grigorenko, K.~Novoselov, T.~Booth, T.~Stauber, N.~Peres,
  and A.~Geim, ``Fine structure constant defines visual transparency of
  graphene,'' {\em Science}, vol.~320, no.~5881, pp.~1308--1308, 2008.

\bibitem{PhysRevLett.102.207209}
F.~Elste, S.~M. Girvin, and A.~A. Clerk, ``Quantum noise interference and
  backaction cooling in cavity nanomechanics,'' {\em Phys. Rev. Lett.},
  vol.~102, p.~207209, May 2009.

\bibitem{PhysRevLett.107.213604}
A.~Xuereb, R.~Schnabel, and K.~Hammerer, ``Dissipative optomechanics in a
  michelson-sagnac interferometer,'' {\em Phys. Rev. Lett.}, vol.~107,
  p.~213604, Nov 2011.

\bibitem{PhysRevApplied.3.024004}
R.~M. Cole, G.~A. Brawley, V.~P. Adiga, R.~De~Alba, J.~M. Parpia, B.~Ilic,
  H.~G. Craighead, and W.~P. Bowen, ``Evanescent-field optical readout of
  graphene mechanical motion at room temperature,'' {\em Phys. Rev. Applied},
  vol.~3, p.~024004, Feb 2015.

\bibitem{sample}
{Graphene Platform Inc., CVD Single layer Graphene Transferred on TEM grid 2000
  mesh}.

\bibitem{cheon2014reliably}
S.~Cheon, K.~D. Kihm, H.~goo Kim, G.~Lim, J.~S. Park, and J.~S. Lee, ``How to
  reliably determine the complex refractive index (ri) of graphene by using two
  independent measurement constraints,'' {\em Scientific Reports}, vol.~4,
  2014.

\bibitem{PhysRevX.4.021052}
M.~Wu, A.~C. Hryciw, C.~Healey, D.~P. Lake, H.~Jayakumar, M.~R. Freeman, J.~P.
  Davis, and P.~E. Barclay, ``Dissipative and dispersive optomechanics in a
  nanocavity torque sensor,'' {\em Phys. Rev. X}, vol.~4, p.~021052, Jun 2014.

\bibitem{PhysRevLett.105.097401}
E.~Hendry, P.~J. Hale, J.~Moger, A.~K. Savchenko, and S.~A. Mikhailov,
  ``Coherent nonlinear optical response of graphene,'' {\em Phys. Rev. Lett.},
  vol.~105, p.~097401, Aug 2010.

\bibitem{brenneis2015ultrafast}
A.~Brenneis, L.~Gaudreau, M.~Seifert, H.~Karl, M.~S. Brandt, H.~Huebl, J.~A.
  Garrido, F.~H. Koppens, and A.~W. Holleitner, ``Ultrafast electronic readout
  of diamond nitrogen-vacancy centres coupled to graphene,'' {\em Nature
  nanotechnology}, vol.~10, no.~2, pp.~135--139, 2015.

\bibitem{PhysRevB.79.165428}
B.~Krauss, T.~Lohmann, D.-H. Chae, M.~Haluska, K.~von Klitzing, and J.~H. Smet,
  ``Laser-induced disassembly of a graphene single crystal into a
  nanocrystalline network,'' {\em Phys. Rev. B}, vol.~79, p.~165428, Apr 2009.

\bibitem{Roberts2011}
A.~Roberts, D.~Cormode, C.~Reynolds, T.~Newhouse-Illige, B.~J. LeRoy, and A.~S.
  Sandhu, ``Response of graphene to femtosecond high-intensity laser
  irradiation,'' {\em Applied Physics Letters}, vol.~99, no.~5, 2011.

\bibitem{graphenefibertip}
Q.~Bao, H.~Zhang, Y.~Wang, Z.~Ni, Y.~Yan, Z.~X. Shen, K.~P. Loh, and D.~Y.
  Tang, ``Atomic-layer graphene as a saturable absorber for ultrafast pulsed
  lasers,'' {\em Advanced Functional Materials}, vol.~19, no.~19,
  pp.~3077--3083, 2009.

\bibitem{xiao2014dissipative}
L.-D. Xiao, Y.-F. Shen, Y.-C. Liu, M.-Y. Yan, and Y.-F. Xiao, ``Dissipative
  optomechanics of a single-layer graphene in a microcavity,'' {\em arXiv:1411.2202}, 2014.

\bibitem{Wu2014}
M.~Wu, A.~C. Hryciw, C.~Healey, D.~P. Lake, H.~Jayakumar, M.~R. Freeman, J.~P.
  Davis, and P.~E. Barclay, ``Dissipative and dispersive optomechanics in a
  nanocavity torque sensor,'' {\em Phys. Rev. X}, vol.~4, p.~021052, Jun 2014.

\end{thebibliography}
\end{document}